\documentclass[10pt,journal,cspaper,compsoc]{journeevisu}

\usepackage[pdftex]{graphicx} 

\usepackage{t1enc}

\usepackage{cite}

\usepackage{float}
\usepackage{microtype}
\usepackage{hyperref}

\usepackage{orcidlink}

\TitreVisu{Visualization in Motion in Video Games for Different Types of Data}{Visualisation Localisée en Mouvement dans les Jeux Vidéo pour Différents Types de Données}

\ShortBib{Federica Bucchieri \MakeLowercase{\textit{et al.}}: Visualization in Motion in Video games for different types of data}

\author{ Federica Bucchieri \orcidlink{0009-0009-6398-0660}, Lijie Yao \orcidlink{0000-0002-4208-5140}, and Petra Isenberg \orcidlink{0000-0002-2948-6417} 
\IEEEcompsocitemizethanks{
\IEEEcompsocthanksitem Federica Bucchieri | Lijie Yao | Petra Isenberg: Universit{\'e} Paris-Saclay, CNRS, Inria, LISN, Orsay, France
    \protect\\E-mail: federica.bucchieri@inria.fr, yaolijie0219@gmail.com, petra.isenberg@inria.fr. 
}
}

\IEEEcompsoctitleabstractindextext{
\begin{abstract}
   We contribute an analysis of situated visualizations in motion in video games for different types of data, with a focus on quantitative and categorical data representations. 
   Video games convey a lot of data to players, to help them succeed in the game. These visualization frequently move across the screen due to camera changes or because the game elements themselves move. Our ultimate goal is to understand how motion factors affect visualization readability in video games and subsequently the players' performance in the game. We started our work by surveying the characteristics of how motion currently influences which kind of data representations in video games.
   We conducted a systematic review of 160 visualizations in motion in video games and extracted patterns and considerations regarding where, what, and how visualizations currently exhibit motion factors in video games. 
\end{abstract}
}

\begin{document}

\maketitle

\section{Introduction}
While playing a video game players generate dynamic datasets about character movements, health, resources, and many other game elements. This data is often visualized to help players succeed and make decisions about their actions in the game. Frequently, those data visualizations are in motion on the screen due to camera changes or because the visualizations are attached to moving game elements. In order to use the data shown in these visualizations, players need to read them at a glance while they focus on the game's primary task. Our research aims to explore how contextual factors in video games affect the readability of visualizations in motion. We conducted a systematic review surveying the characteristics of visualizations in motion in video games. We collected 160 visualizations from 50 video games. Moreover, we analyzed how different types of data are represented in games. Specifically, we focused on quantitative and categorical data by analyzing two prominent examples of the information displayed in video games: character's health and the type of specific game elements.

\section{Related Work}
Our work focuses on \emph{situated visualizations in motion for video games}. Situated visualizations are data representations located near a data referent \cite{Willett:2017:EDR}. A data referent is a person, object, or location to which the visualized data refers. Willet et al.\ studied the relationship between visual representations and data referents, formalizing the difference between situated and embedded data visualizations. Their work is based on White's \cite{White:2009:IPTforSituatedVis} investigation of presentation and interaction techniques. White proposed a theoretical framework for situated visualization, presenting best practices for the domain. Our research is related to situated visualization since we are considering game element visualizations displayed in close connection with the element itself. The notion of \emph{visualization in motion} was firstly introduced by Yao et al.\ \cite{Yao:2020:SituatedVisinMotion} in their design space. They explored motion relationships between viewers and visualizations. They analyzed motion factors and contextual factors that may affect the readability of the visualizations. In our previous \cite{Bucchieri:2022:SituatedVisinMotionforGames} and current work, we have narrowed down Yao et al.'s research scope by focusing on static viewers and moving visualizations, in the context of video games visualizations. Our work is naturally related to previous work in video game visualization techniques. In 2008, Zammitto \cite{Zamitto:2008:VisTechInVideoGames} analyzed how video games present essential information to the players and if visualization techniques were used. Zammitto presented visual representations such as health bars, silhouettes, HUDs, and mini-maps. Bowman et al.\ \cite{Bowman:2012:TowardVisforGames} conducted a systematic review of existing games incorporating visualization to build a framework. They proposed a design space to classify any video game visualizations according to five categories: primary purpose, target audience, temporal usage, visual complexity, and immersion/integration. Our work relates to Peacocke et al.'s \cite{Peacocke:2018:EmpiricalComparisonFPS} evaluation of player's performances about different types of data represented in video games with different types of displays. The authors showed that different visual representations are suited for specific and diverse types of information. Our study differs from previous studies in that it focuses on visualizations impacted by motion factors in the context of game-specific elements.

\vspace{-10pt}
\section{Systematic Review of Visualization in Motion in Video Games}
Our work is based on a systematic review of visualizations in motion for video games \cite{Bucchieri:2022:SituatedVisinMotionforGames}. We used a commercial ranking website called Metacritic\cite{Metacritic} to cover a diverse range of video games. We collected 160 visualizations in motion by surveying 50 video games from 17 genres. We selected the top 3 games from 2011 to 2022 sorted by the Metacritic relevance score for each genre. We categorized those 160 visualizations according to several dimensions related to situated visualization and motion factors. In our previous EuroVis poster we presented an analysis of five of those dimensions, while here we concentrate on how different types of data are represented in video games.

\vspace{-10pt}
\section{Visualization in motion in video games for different types of data}
Video games produce a wide range of types of data, defined by the values they can express. 67 out of the 160 visualizations we analyzed represented \emph{quantitative data}. This data concerned health points of a character, the number of resources crafted, or the distance from a location.
\emph{Categorical data} were the second most common category, with 64 out of 160 representatives. Categorical data concerned team identification and resource type.
Furthermore, 49 out of the 160 visualizations showed \emph{spatial data} of the data referent's position. 
Meanwhile, \emph{ordered data} were the least frequent with 8 out of 160 representatives and were only present in racing games. In the next two sections we focus on the two most frequent quantitative and categorical data displays: character health and game element types.

\vspace{-10pt}
\subsection{Character's health}
\begin{figure}[H]
  \centering
  \includegraphics[width=.323\linewidth]{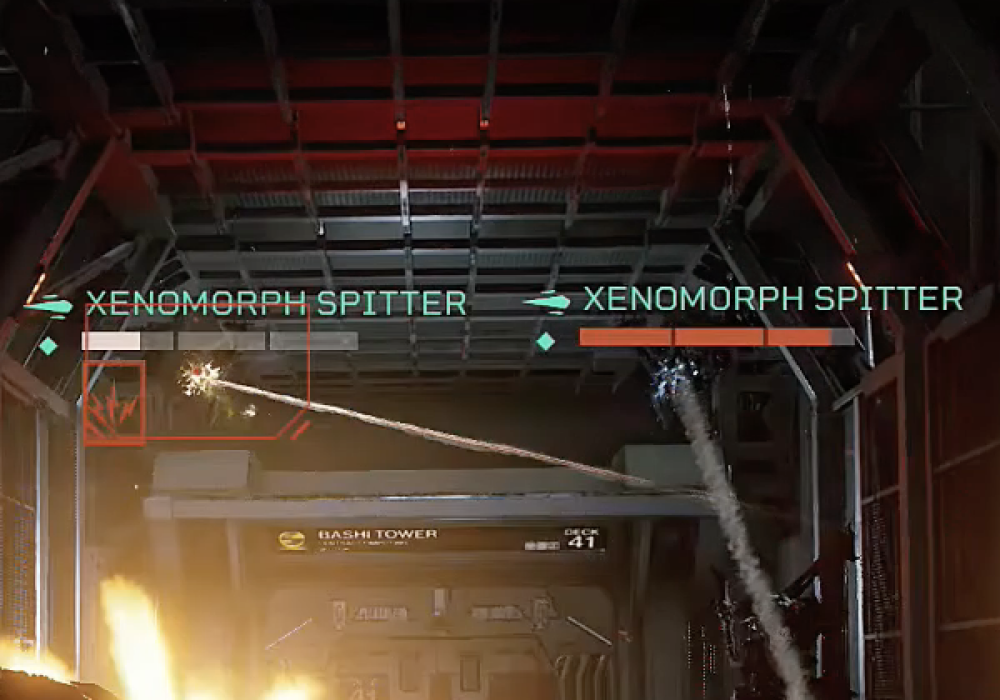}
  \hfill
  \includegraphics[width=.323\linewidth]{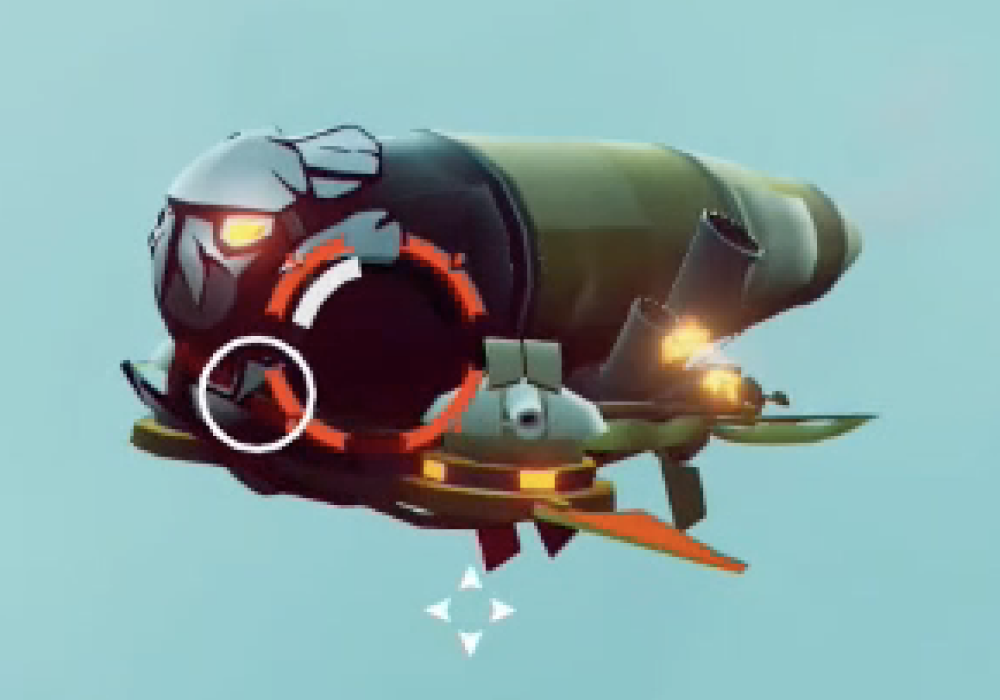}
\hfill
  \includegraphics[width=.323\linewidth]{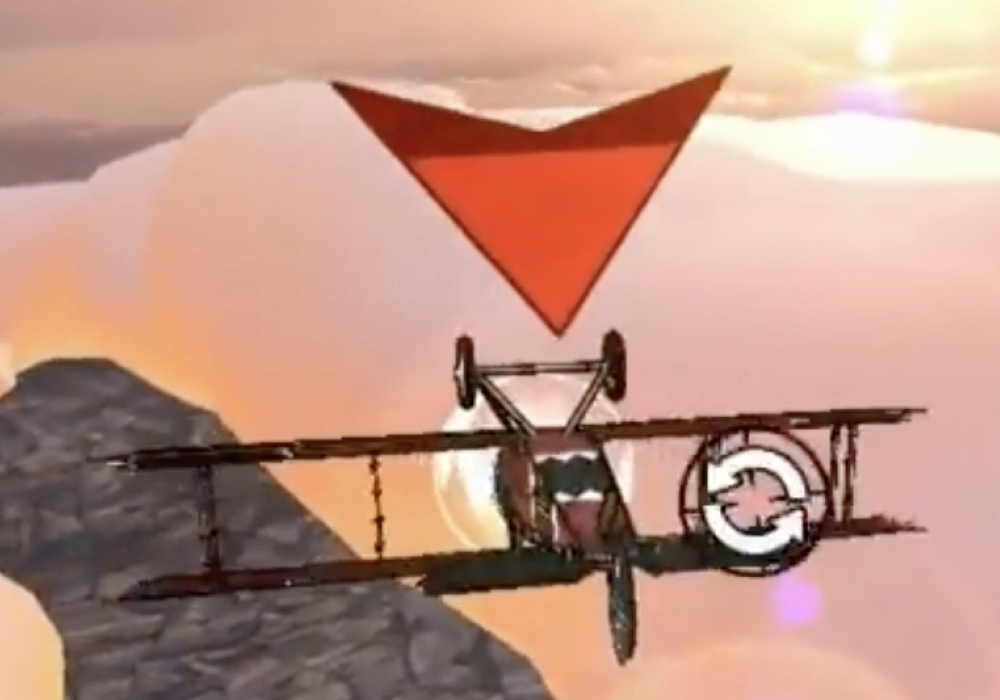}
  \caption{\label{fig:ex2}%
           Different types of character's health's visualizations. Left: Bar chart in \emph{Aliens: Fireteam Elite}. Middle: Radial bar chart in \emph{The Falconeer}. Right: Pictorial fraction chart in \emph{Skies of Fury DX}.
          }
\end{figure}

Out of 67 \emph{quantitative data} visualizations, 24 represented the  character's health. A typical way to represent health were horizontal bar charts (18/24 visualizations; see Fig.\ \ref{fig:ex2}: Left). Only 3 out of 24 representatives were radial bar charts (see Fig.\ \ref{fig:ex2}: Middle), and the remaining three were a pie chart, a label with numbers, and a pictorial fraction chart (see Fig.\ \ref{fig:ex2}: Right). Health was sometimes represented with single-color horizontal bar charts that encoded data only by bar length. Some examples showed a dual encoding with the use of color, usually in gradients from green (healthy) to red (critical). As stated by Zammitto \cite{Zamitto:2008:VisTechInVideoGames}, to be color-blind friendly, applying length as the unique method is already sufficient. 16/24 visualizations in our collection had an opaque background, while the remaining 8 had a transparent background. Having a transparent background also meant that these bars lacked a reference frame that could help to judge the length of the bar more precisely \cite{ClevelandMcGill:1984:GraphicalPerception}.

\vspace{-5pt}

\vspace{-20pt}
\subsection{Game element types}
\begin{figure}[h]
  \centering
  \includegraphics[width=.323\linewidth]{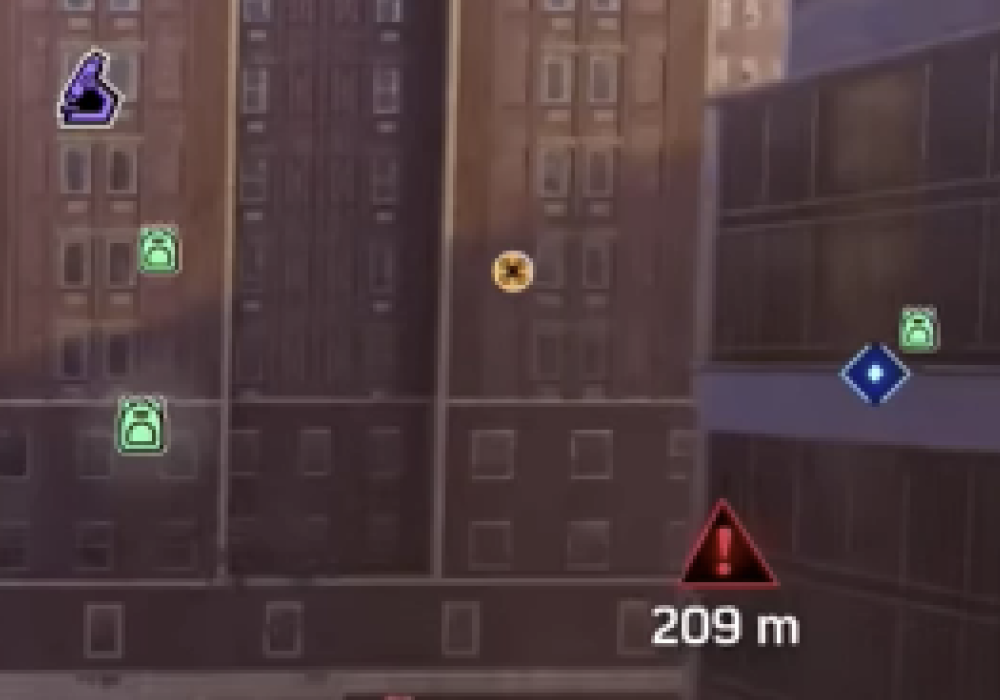}
  \hfill
  \includegraphics[width=.323\linewidth]{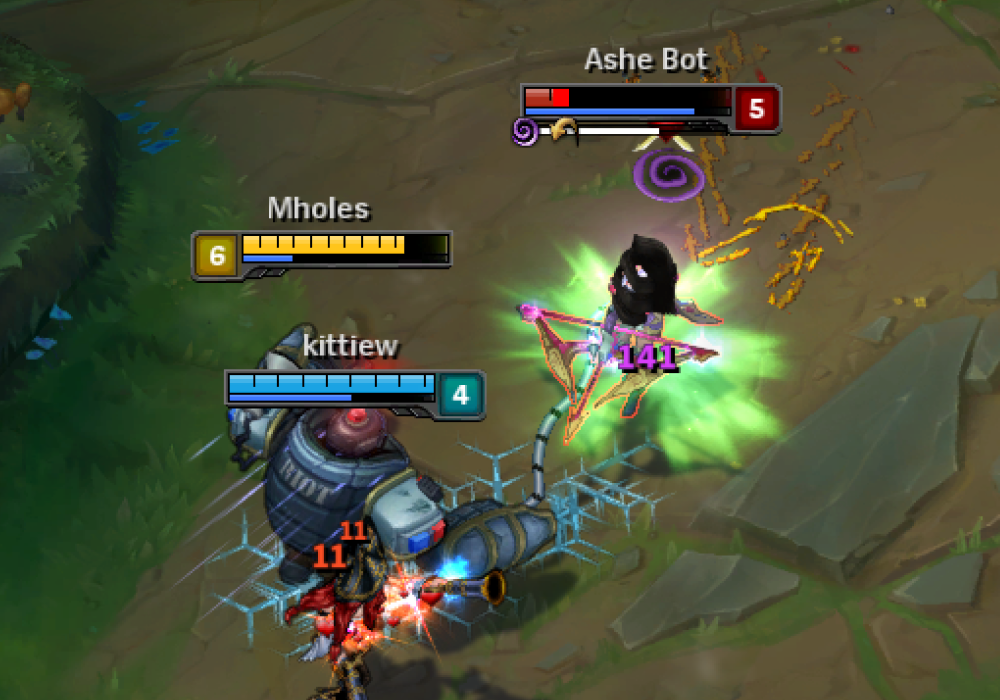}
\hfill
  \includegraphics[width=.323\linewidth]{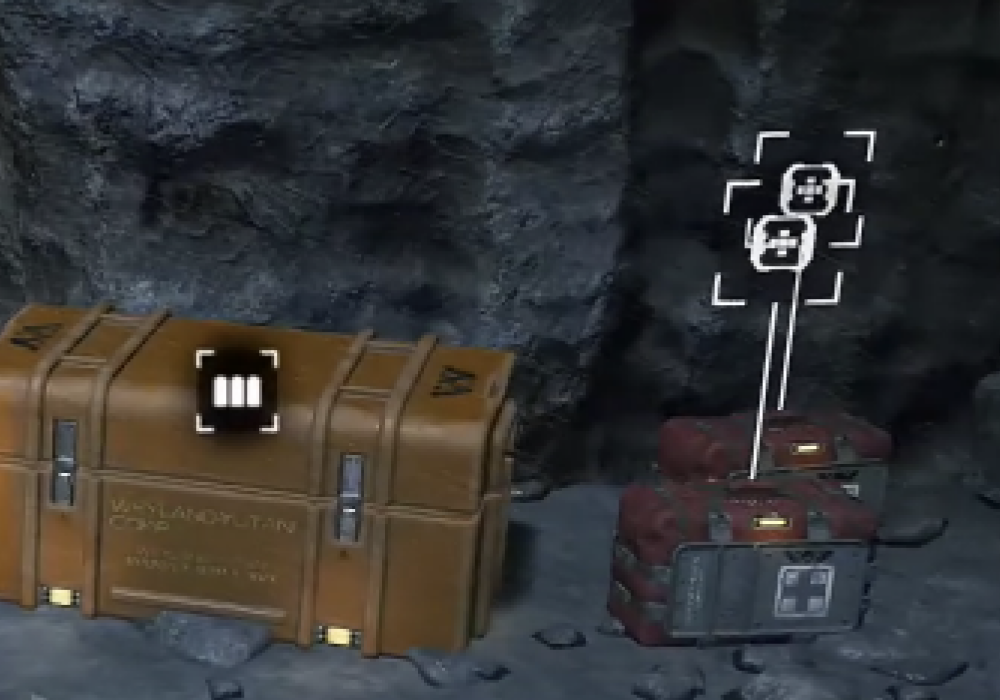}
  \caption{\label{fig:ex3}%
           Different game element's type visualizations. Left: missions' icons in \emph{Spider-Man}. Middle: color coded health bar charts in \emph{League of Legends}. Right: Resource's type signs in \emph{Aliens: Fireteam Elite}.
          }
\end{figure}

To succeed in certain games, finding the right resource or correctly identifying enemies is important. This type of information is usually represented by \emph{categorical data}. We found 26/160 visualizations that represented the type of game elements, such as interactive objects and characters. Signs were the most common visual representation, with 13/26 visualizations. Signs used shape and color encodings to distinguish different categories of game elements, such as mission targets (see Fig.\ \ref{fig:ex3}: Left). Those encodings aimed to help players differentiate between other players, objects, and objectives \cite{Karlsson:2016:information}. Another prevalent method to identify character types is the use of color in the character's health bar charts (see Fig.\ \ref{fig:ex3}: Middle). In fact, when categorizing the 26 visualizations found by the type of encoding used, it's possible to notice that the majority used color-only with 12 out of 26 representatives, while 8 out of 26 visualizations used both shape and color, and 5 used only shape (see Fig.\ \ref{fig:ex3}: Right). The last representative (1/26) displayed information by using text.

\vspace{-5pt}

\vspace{-15pt}
\section{Discussion and Future Work}
Our results show that quantitative and categorical data are the most recurrent types of data visualized in video games.
According to our sample, the prevalent encoding used for quantitative data is length. Instead, color is mostly used for categorical data representations. Character's health (quantitative data) is predominantly represented by bar charts while, game element types (categorical data) by using signs. We are now designing an empirical study to evaluate the readability of different visual representations under motion in the context of video games.

\vspace{-5pt}
\section{Acknowledgement }
This work was partly supported by the Agence Nationale de la Recherche (ANR), grant number ANR-19-CE33-0012.

\bibliographystyle{abbrv}
\bibliography{abstract}

\end{document}